\documentclass[12pt]{article}

\usepackage[utf8]{inputenc}               
\usepackage[T1]{fontenc}                  
\usepackage[font=small]{caption}          
\usepackage{amsmath, amsthm}              
\usepackage{amsfonts}                     
\usepackage{mathtools}                    
\usepackage{bm, bbm}                      
\usepackage{enumitem}                     
\usepackage{url}                          
\usepackage{graphicx}                     
\usepackage{float}                        
\usepackage{placeins}                     
\usepackage{xcolor}                       
\usepackage{setspace}                     
\usepackage{orcidlink}                    

\usepackage[authoryear,round,comma]{natbib}  

\usepackage{hyperref}                     
\hypersetup{colorlinks=true,citecolor=black,linkcolor=black,urlcolor=black}

\usepackage{authblk}                      

\usepackage{geometry}                     
\usepackage{appendix}                     

\newcommand{\beginappendix}{              
    \renewcommand{\thesection}{\Alph{section}}
    \renewcommand{\thesubsection}{\Alph{section}.\arabic{subsection}}
}

\usepackage{listings}                     
\definecolor{DarkGreen}{RGB}{1,50,32}
\newcommand{\SUM}[2]{\sum_{#1}^{#2}}

\newcommand{\dnorm}[3]{\phi \left( #1 ; #2, #3 \right)}

\def\hy{\hat{y}}
\def\hs{\hat{s}}
\def\map{\mathrm{map}}
\def\rmap{\mathrm{rmap}}
\def\pow{\mathrm{pow}}
\def\rpow{\mathrm{rpow}}
\def\pr{\mathrm{pr}}

\begin{document}

\title{\bf On regional treatment effect assessment using robust MAP priors}

\author[1]{Xin Zhang\orcidlink{0000-0001-9894-2634}\thanks{Corresponding author. Email: \href{xin.zhang6@pfizer.com}{xin.zhang6@pfizer.com}.}}
\author[1]{Hui Zhang\thanks{Email: \href{hui.zhang2@pfizer.com}{hui.zhang2@pfizer.com}.}}
\author[1]{Satrajit Roychoudhury\orcidlink{0000-0003-4001-3036}\thanks{Email: \href{satrajit.roychoudhury@pfizer.com}{satrajit.roychoudhury@pfizer.com}.}}

\affil[1]{Data Sciences and Analytics, Pfizer Inc.}

\date{}

\maketitle

\begin{abstract}

Bayesian dynamic borrowing has become an increasingly important tool for evaluating the consistency of regional treatment effects which is a key requirement for local regulatory approval of a new drug. It helps increase the precision of regional treatment effect estimate when regional and global data are similar, while guarding against potential bias when they differ. In practice, the two-component mixture prior, of which one mixture component utilizes the power prior to incorporate external data, is widely used. It allows convenient prior specification, analytical posterior computation, and fast evaluation of operating characteristics. Though the robust meta-analytical-predictive (MAP) prior is broadly used with multiple external data sources, it remains underutilized for regional treatment effect assessment (typically only one external data source is available) due to its inherit complexity in prior specification and posterior computation. In this article, we illustrate the applicability of the robust MAP prior in the regional treatment effect assessment by developing a closed-form approximation for its posterior distribution while leveraging its relationship with the power prior. The proposed methodology substantially reduces the computational burden of identifying prior parameters for desired operating characteristics. Moreover, we have demonstrated that the MAP prior is an attractive choice to construct the informative component of the mixture prior compared to the power prior. The advantage can be explained through a Bayesian hypothesis testing perspective. Using a real-world example, we illustrate how our proposed method enables efficient and transparent development of a Bayesian dynamic borrowing design to show regional consistency.

    
\end{abstract}


\section{Introduction} 

In modern global drug development, evaluating the consistency of treatment effects between regional and overall populations is critical for local regulatory approval, as discussed in many literature \citep[][among others]{shih2001clinical,hung2010consideration,yusuf2016interpreting,quan2017example,li2020lessons}, and thereafter highlighted in the regulatory guideline \citep{ich2017}. Regional data required by local regulators for drug approval typically come from standalone bridging studies conducted separately from global trials or regional subgroups within multiregional clinical trials, when such regions are included in the global program. Consistency assessment is often challenging because the sample size of the target population is rarely sufficiently powered. This leads to potential unstable estimates and unreliable evaluation of regional treatment effects. 

Bayesian dynamic borrowing \citep{schmidli2014robust,best2021assessing} has become an increasingly popular framework for incorporating external data into statistical analyses. In contrast to other informative prior approaches, such as power priors \citep{ibrahim2000power}, this approach adaptively determines the extent of borrowing based on the similarity between target and external data. Since regional data are typically accompanied with data from other regions, Bayesian dynamical borrowing offers an appealing strategy to strengthen the regional treatment effect assessment and is explicitly referenced in regulatory guidelines \citep{ich2017}.  

When using Bayesian dynamical borrowing for the regional treatment effect assessment, informative priors are typically specified as two-component mixture priors, combining a power prior with a weakly informative prior \citep{hsiao2007use,edwards2024using}.  The power prior is a convenient tool for incorporating information from a single external data source. However, relying solely on power priors risks biasing the results toward external data, affecting the consistency assessment \citep{liu2002bayesian}. Mixing with weakly informative priors allows the model to \textit{dynamically} decide on borrowing from external data. Thus, those mixtures reduce bias when target and external data are dislike. We refer to this prior as the \textit{robust power prior}. 
  
Alternatively, \textit{robust meta-analytical-predictive (MAP) priors} \citep{schmidli2014robust} are also widely used for Bayesian dynamic borrowing. However, their application to regional treatment effect assessment remains uncommon, since that this approach is most appropriate when multiple external data sources are available. The robust MAP prior is also a two-component mixture, while the MAP prior derived using external data \citep{neuenschwander2010summarizing} serves as the informative component. The MAP prior assumes that the true treatment effects across all data sources, including the target data, are exchangeable, thereby enabling information sharing. In practice, the MAP prior with a single external data source is tricky to implement as its posterior distribution is sensitive to the prior specification of the heterogeneity parameter \citep{rover2020dynamically}.  

When applying Bayesian dynamic borrowing in practice, prior parameter choices are typically guided by study design requirements to achieve specific type I error rates and power. This creates an additional obstacle for using robust MAP priors, as evaluating their operating characteristics is time-consuming. Additionally, discounting parameters in power priors provide an interpretable measure of the magnitude of external information to be incorporated in target data analyses, whereas this is less clear for MAP priors. 

In this article, we demonstrate that the robust MAP prior is an effective approach for assessing regional treatment effects. Specifically, we derive a closed-form approximation of the posterior distribution under the robust MAP prior by exploiting its connection to the power prior. This approximation substantially reduces the computational burden, enabling rapid identification of prior parameters that meet the desired operating characteristics--typically within minutes. Moreover, the closed-form solution produces a secondary output: a visual representation of the degree to which external information influences the analysis. Finally, we show that, from a Bayesian hypothesis-testing perspective, the MAP prior model offers distinct advantages over the power prior model.

The remainder of this article is organized as follows. We begin with the motivating example (Section~\ref{sec:ex}), followed by an introduction to two mixture priors for Bayesian dynamic borrowing (Section~\ref{sec:pre}). Section~\ref{sec:method} presents the main results. Using the motivating example, we demonstrate in Section~\ref{sec:app} how these formulae enable the efficient and transparent development of a Bayesian dynamic borrowing design. Section~\ref{sec:disc} concludes our article with a discussion of possible extensions. All technical derivations can be found in Appendix~\ref{append:rpow-posterior}-\ref{append:proof-rmap-posterior-approx}, and the R code for replication purposes is provided in Appendix~\ref{append:code}.  

\section{Motivating Examples} \label{sec:ex}

Our motivating example is a real-world case study presented in \citet{edwards2024using} on planning and evaluating a Bayesian dynamic borrowing design of a bridging study to support the registration of a new medicine with the Center for Drug Evaluation (CDE) in China. A global study had been conducted to support worldwide approval of the drug, however, it did not recruit any Chinese patients. CDE had required a separate bridging study for local registration. To strengthen the evidence for treatment effects in the Chinese subpopulation, the robust power prior approach was applied in this bridging study design to incorporate information from the global study.

The informative component of this mixture prior is a power prior developed based on the aggregate data from the global study without any further discounting (i.e., full borrowing). The success criterion is based on the posterior probability that there is no effect in the Chinese subpopulation is no greater than 5\%. Based on the bridging concept, scientific considerations initially favored assigning a substantial prior mixture weight to the informative component. Nevertheless, to maintain the $\alpha$-level at $0.2$ under the predefined success criterion, the weight was constrained to $0.3$, which is counterintuitive as it indicates that the prior belief of a consistent regional treatment effect is implausible (provided that a mixture weigh of 0.5 indicates a neutral position). To resolve issue, one may adopt a robust power prior incorporating discounted external information or alternatively employ a robust MAP prior. In Section~\ref{sec:app}, we will illustrate how to identify the corresponding prior parameters for given operating characteristics, with the method proposed in Section~\ref{sec:method}.

\section{Preliminaries: Bayesian Dynamic Borrowing} \label{sec:pre}

In this article, we consider Bayesian dynamic borrowing with a single (trial-based) external data source, of which those subjects are defined as external group. We further define the subjects from the region of interest as the target group. The target group can be participants in a bridging study or regional subgroup within a multiregional trial. The goal is to leverage data of the external group for the consistency assessment of the target group using Bayesian dynamic borrowing.  

The following notation is adopted for use throughout the article. Let $( n_\ast, \hy_\ast, \hs_\ast, \theta_\ast )$ and $( n_1, \hy_1, \hs_1, \theta_1 )$ be the sample sizes, the estimated treatment effects, the estimated standard errors, and the true treatment effects for the target and the external group, respectively. In addition, $\sigma_\ast^2$ and $\sigma_1^2$ are the within-group variances for the target and the external group, respectively. They could be the same, equaling the assumed variance used in the sample size calculation of the global study (i.e. $\sigma_\ast^2 = \sigma_1^2 = \sigma_0^2$). Without loss of generality, we let the null value for $\theta_\ast$ and $\theta_1$ being zero and $\theta_1, \theta_\ast > 0$ indicate that the investigating treatment is beneficial.  

\subsection{Meta-analytical predictive (MAP) and Power priors} \label{sec:pre-map}

The MAP prior \citep{neuenschwander2010summarizing} is one of the most commonly used methods to develop informative priors from external data. It explicitly relates $\theta_\ast$ and $\theta_1$ through a hierarchical model by assuming exchangeability. This assumption implies that there is no prior knowledge suggesting that $\theta_\ast$ is superior or inferior to $\theta_1$. Specifically, $\theta_1, \theta_\ast \sim \dnorm{\cdot}{\mu}{\tau^2}$, where $\phi(\cdot)$ denotes the density function of a normal distribution. The common mean $\mu$, and the standard deviation $\tau$ represent the overall mean and between-group heterogeneity. With additional priors for both $\mu$ and $\tau$, written as $\mu \sim \dnorm{\cdot}{m}{v}$ and $\tau \sim f_0 ( \cdot; \nu )$, the hierarchical model allows estimation of $\theta_\ast$ by borrowing information from $\hy_1$, while still acknowledging that $\theta_\ast$ and $\theta_1$ need not be identical. This hierarchical model leads to the MAP prior, which is expressed as the predictive distribution of $\theta_\ast$, derived from the posterior distribution of $( \mu, \tau )$ given $\hy_1$. We denote the MAP prior by $p_\map ( \theta_\ast \vert \hy_1, \hs_1^2, f_0 )$, omitting its dependence on $(m,v)$ since they are generally not of interests.  

The MAP prior is initially developed to borrow information from multiple external data sources, while, when only a single external data source is available, the power prior \citep{ibrahim2000power} is commonly used. This approach incorporate external data in the likelihood with a discounting factor $\lambda \in [0,1]$. This parameter $\lambda$ governs the extent of the information borrowed from $\hy_1$ in the power prior: $\lambda = 0$ indicates no borrowing while $\lambda = 1$ indicates complete pooling. The amount of borrowed information in the power prior can be quantified by the prior effective sample size $\lambda n_1$ \citep{morita2008determining,neuenschwander2016use}. 

The MAP and power priors are mathematically related \citep{chen2006relationship,neuenschwander2016use}. Assuming that $\hs_1^2 = \sigma_1^2 / n_1$, $m = 0$, $v \to \infty$, and $f_0 ( \cdot; \nu )$ is a point mass at $\tau > 0$, the MAP prior can be written as follows  \citep[][Theorem 2.2]{chen2006relationship}
\begin{equation} \label{eq:tau2lambda}
p_\map ( \theta_\ast \vert \hy_1, \hs_1^2, f_0 ) = \dnorm{\theta_\ast}{\hy_1}{ \hs_1^2 / \lambda } \text{, where }
\lambda = \frac{1}{ 2n_1\tau^2/\sigma_1^2 + 1 }
\end{equation}
The above relationship implies that the between-group heterogeneity can be translated into the discounting factor of the power prior. This relationship between $\tau^2$ and $\lambda$ also suggests that the specification of $f_0 ( \tau; \nu )$ is crucial when there is only one external data source.   

Furthermore, as \citet{neuenschwander2016use} pointed out, the underlying hierarchical model for the MAP prior also implies the commensurate prior \citep{hobbs2012commensurate} when there is only one external data sources.  Again, assuming $m = 0$ and $v \to \infty$ and following the formulation of the commensurate prior, the joint distribution of $ ( \theta_\ast, \theta_1 ) $ given $\hy_\ast$ is proportional to $\dnorm{\hy_1}{\theta_1}{\hs_1^2} \cdot \int\! \dnorm{\theta_\ast}{\theta_1}{2\tau^2} f_0 ( \tau; \nu ) \, \mathrm{d} \tau$.  This joint distribution suggests that the bias of $\theta_\ast$ from $\theta_1$ is determined by $\tau$, and thus $f_0 ( \tau; \nu )$ captures the uncertainty in such bias.  \citet{rover2020dynamically} suggests using weakly informative priors for $f_0 ( \tau; \nu )$, and \citet{rover2021on} offer comprehensive guidance on specifying such priors using the unit information standard deviation \citep{rover2021on}, which reflects an observational unit's contribution to the study's likelihood.  However, these methods do not guarantee that the resulting design will achieve the desired type I error rate and power.

\subsection{Mixture priors} \label{sec:pre-bdb}

In practice, the MAP or power prior is usually combined with a weakly informative prior, through a mixture model, to handle the ``prior–data conflict'' \citep{schmidli2014robust,best2021assessing}. A ``prior–data conflict'' refers to a scenario in which external and target data for the treatment effects are not aligned. When using the MAP prior as the informative component, the mixture prior is expressed as 
\[
p_\rmap ( \theta_\ast \vert \hy_1, \hs_1^2, f_0, w_0 ) = ( 1 - w_0 ) \cdot \dnorm{\theta_\ast}{0}{\sigma_0^2} + w_0 \cdot p_\map ( \theta_\ast \vert \hy_1, \hs_1^2, f_0),
\]
where $w_0\in[0,1]$ is the mixture weight for the informative component. This prior is known as the robust MAP prior \citep{schmidli2014robust}. The second component can be replaced with the power prior, which we refer to this alternative mixture prior as the robust power prior.  

In the mixture, the first component is a weakly informative prior (e.g., unit information prior), centered on the null treatment effect, which represents no borrowing from $\hat{y}_1$. The mixture weight $w_0$ represents the prior probability that $\theta_\ast$ and $\theta_1$ are related through the informative prior model. Both $w_0$ and $f_0 ( \tau; \nu )$ (or $\lambda$ if using the power prior as the informative component) govern the influence of $\hy_1$ on the posterior distribution of $\theta_\ast$.  

Moreover, we denote $\pi_{\rmap} ( \theta_\ast \vert \hy_\ast, \hs_\ast^2, \hy_1, \hs_1^2, f_0, w_0 )$ and $\pi_{\rpow} ( \theta_\ast \vert \hy_\ast, \hs_\ast^2, \hy_1, \hs_1^2, \lambda, w_0 )$ as the posterior distributions for the robust MAP and robust power priors, respectively. Evaluation of $\pi_{\rmap} ( \cdot )$ often requires intensive computation, while $\pi_{\rpow}$ has a simple closed-form expression (Appendix~\ref{append:rpow-posterior}). This makes the latter more practically appealing for real-world applications \citep{hsiao2007use,koppschneider2020power}, especially for evaluating operating characteristics in study planning. 

\subsection{Operating characteristics} \label{rem:pre-bdb-oc}

In this setting, the investigating treatment is considered effective in a region if the posterior probability crosses certain predefined threshold, for example, $\pr ( \theta_\ast > 0 \vert \hy_\ast, \hs_\ast^2, \hy_0, \hs_1^2 ) > 95 \%$.  In \citet{weber2021applying}, a two-step method is applied to calculate operating characteristics for such designs. Let $y_c$ be the critical decision boundary, which is the minimal observed regional treatment effect size required for the posterior probability reaching the predefined threshold. The first step is to determine $y_c$ via a grid search. Next, the type I error rate and power (at the alternative $y_a$) are computed as 
\begin{eqnarray*}
\text{Type-I error} &=& 1 - \Phi \left( \frac{ \sqrt{n} \{ y_c - 0 \} }{ 2\sigma_0 } \right) \\ 
\text{Power} &=& - \Phi \left( \frac{ \sqrt{n} \{ y_c - y_a \} }{ 2\sigma_0 } \right).
\end{eqnarray*}
 
Since $\pi_{\rpow} ( \cdot )$ has a closed-form expression, $y_c$ can be identified without simulation, and thus both the type I error and power can be efficiently computed. For $\pi_{\rmap} ( \cdot )$, \citet{weber2021applying} propose approximating the predictive distribution with a mixture of conjugate distributions. It enables numerical integration to compute the type I error rate and power. However, substantial computational effort is needed to get a sufficient number of samples from the predictive distribution and a good approximation for $\pi_{\rmap} ( \cdot )$, and thus its operating characteristics evaluation is still computationally intensive. 

\section{Methodology}  \label{sec:method}

In this section, we develop a closed-form estimator to approximate the posterior distribution for the robust MAP prior (Section~\ref{sec:method-posterior}). We also demonstrate that the robust MAP prior offers distinct advantages over the robust power prior from a Bayesian hypothesis testing perspective (Section~\ref{sec:method-bf}). 

\subsection{Posterior distributions} \label{sec:method-posterior}
 
Our estimator to approximate $\pi_{\rmap} ( \cdot )$ is motivated by the closed-form formula for $\pi_{\rpow} ( \cdot )$ (Appendix~\ref{append:rpow-posterior}) and the relationship between $\tau$ and $\lambda$ as described in \eqref{eq:tau2lambda}.  

First of all, we express $p_\map ( \cdot )$ as (assuming $m \to 0$ and $v \to \infty$)
\begin{equation} \label{eq:map-powmix}
p_\map ( \theta_\ast \vert \hy_1, \hs_1^2, f_0 ) = \int \! \dnorm{\theta_\ast}{\hy_1}{ \hs_1^2 + 2\tau^2 } f_0 ( \tau; \nu ) \, \mathrm{d} \tau.    
\end{equation}
\noindent The proof of \eqref{eq:map-powmix} can be found in Appendix~\ref{append:proof-map-mix-pow}. Since $\dnorm{\theta_\ast}{\hy_1}{ \hs_1^2 + 2\tau^2 }$ corresponds to a power prior with $\lambda = 1 / ( 2\tau^2/\hs_1^2 + 1 )$, \eqref{eq:map-powmix} shows that $p_\map ( \cdot )$ is a continuous mixture of power priors, where the mixing distribution is $f_0 ( \tau; \nu )$. Consequently, $f_0 ( \tau; \nu )$ induces a prior on $\lambda$ and thus on $\lambda n_1$, the effective sample size of the power prior. This formulation provides a natural interpretation of $f_0 ( \tau; \nu )$, and extends the result in \eqref{eq:tau2lambda}, where $f_0 ( \cdot )$ reduces to a point mass.  

In addition, the formulation of $p_\map ( \cdot )$ provided in \eqref{eq:map-powmix} suggests a convenient approach for numerical integration, which is to integrate by $\lambda n_1$. A natural partition of the support, $\lambda n_1 \in ( 0, n_1 ]$, is the following intervals: $(0,1], (1,2], \ldots, (n_1-2,n_1-1], (n_1-1,n_1]$.  This motivates the partition for the support of $\tau \in [ 0, +\infty )$ as follows: 
\begin{multline} \label{eq:partition}
[\tau_{n_1}, \tau_{n_1-1}), \ldots, [\tau_1, \tau_0),\text{ where} \\
\tau_i \coloneq \sqrt{ (  n_1/i - 1 ) \cdot \hs^2_1 / 2 } \text{ for } i = 1, \ldots, n_1, \text{ and } \tau_{0} \coloneq + \infty,
\end{multline}
such that $\tau \in [ \tau_i, \tau_{i-1} ) \implies \lambda n_1 \in ( i - 1, i ]$. Each $\tau_i$ corresponds to a discounting parameter $\lambda_i \coloneqq i/n_1$ in the power prior, with the effective sample size of $\lambda_i n_1 = i$. Applying the partition scheme \eqref{eq:partition} to the integration in \eqref{eq:map-powmix}, we have that the following approximation for the MAP prior:  
\begin{equation} \label{eq:tau-prior}
\begin{split}
\hat{p}_\map ( \theta_\ast \vert \hy_1, \hs_1^2, f_0 ) & = \SUM{i=1}{n_1} h_0(i;\nu) \cdot \dnorm{\theta_\ast}{\hy_1}{ n_1/i \cdot \hs_1^2}, \\
h_0(i;\nu) & = F_0 ( \tau_{i-1}; \nu ) - F_0 ( \tau_i; \nu ), 
\end{split}
\end{equation}
where $F_0 ( \tau; \nu ) $ is the cumulative distribution function of $f_0 ( \tau; \nu )$.  Since $h_0(i;\nu)\geq0$ and $\SUM{i=1}{n_1}h_0(i;\nu)=1$, $\hat{p}_\map (\cdot)$ is a $n_1$-component mixture of power priors. Each $h_0(i;\nu)$ represents the prior probability that $\tau \in [\tau_i, \tau_{i-1})$ or equivalently that $\lambda n_1 \in ( i - 1, i]$. This corresponds to the prior probability of borrowing approximately $i$ subjects from the external data source. Notably, $h_0(1)$ should be small to avoid excessive conservatism, as the last interval $\lambda n_1 \in ( 0, 1 ]$ implies an effective sample size of at most one, which is similar to a weakly informative prior.

Next, using the approximated prior \eqref{eq:tau-prior} we have the following approximated estimator for the posterior $\pi_{\rmap} ( \theta_\ast \vert \hy_\ast, \hs_\ast^2, \hy_1, \hs_1^2, f_0, w_0 )$: 
\begin{multline}
\hat\pi_\rmap ( \theta_\ast \vert \hy_\ast, \hs_\ast^2, \hy_1, \hs_1^2, f_0, w_0 ) = ( 1 - w ) \cdot \dnorm{\theta_\ast}{\hy_\ast}{ \{ \hs_\ast^{-2} + \sigma_0^{-2}\}^{-1} } \\ + w \cdot \SUM{i=1}{n_1} h(i;\nu) \cdot \dnorm{\theta}{ \frac{ \hs_\ast^{-2} \hy_\ast + i / n_1 \cdot \hs_1^{-2} \hy_1 }{ \hs_\ast^{-2} + i / n_1 \cdot \hs_1^{-2} } }{ \frac{ 1 }{ \hs_\ast^{-2} + i / n_1 \cdot \hs_1^{-2} } }, \label{eq:rmap-posterior}
\end{multline}
where
\begin{equation} \label{eq:tau-posterior}
\begin{split}
h(i;\nu) & = \frac{ h_0(i;\nu) \cdot \dnorm{\hy_\ast}{\hy_1}{ \hs_\ast^2 + n_1 / i \cdot \hs^2_1 } }{ \SUM{ j = 1 }{ n_1 } h_0(j;\nu) \cdot \dnorm{\hy_\ast}{\hy_1}{ \hs_\ast^2 + n_1 / j \cdot \hs^2_1 } },  \\
w & = \frac{ B_\nu \cdot w_0 / ( 1 - w_0 ) }{ B_\nu \cdot w_0 / ( 1 - w_0 ) + 1 }, \\
B_\nu & = \SUM{ i = 1 }{ n_1 } h_0(i;\nu) \cdot \frac{ \dnorm{\hy_\ast}{\hy_1}{ \hs_\ast^2 + n_1 / i \cdot \hs_1^2 } }{ \dnorm{\hy_\ast}{0}{ \hs_\ast^2 + \sigma_0^2 } }. 
\end{split}
\end{equation}
\noindent The proof of \eqref{eq:rmap-posterior}--\eqref{eq:tau-posterior} can be found in Appendix~\ref{append:proof-rmap-posterior-approx}.  

The collection $\{ h(i;\nu); i = 1, \ldots, n_1 \}$ represents the posterior distribution of $\tau$. Using the relationship between $\tau$ and $\lambda n_1$, as shown in \eqref{eq:tau2lambda}, this distribution can be interpreted as the posterior probability of incorporating approximately $i$ subjects from the external data source. A comparison of $\{ h(i;\nu) \}$ with $\{ h_0(i;\nu) \}$ provides a graphical assessment of the extent to which external information influences the analysis.

\subsection{Bayesian hypothesis testing} \label{sec:method-bf}

The mixture prior, either $p_{\rpow} (\cdot)$ or $p_{\rmap} (\cdot)$, represents a weighted average of two prior models \citep{hoeting1999bayesian,rover2019model}, under two distinct assumptions. Specifically, the weakly informative prior assumes that $\theta_\ast$ and $\theta_1$ are unrelated, while the informative component assumes these are related. The resulting $\pi_\rmap(\cdot)$ and $\pi_\rpow(\cdot)$ combine inference from both assumptions, with the posterior mixture weight ($w$) indicating which assumption is supported by the observed data ($\hy_\ast$).  In particular, $w$ is determined by the corresponding Bayes factor $B_\lambda$ and $B_\nu$, respectively; see Appendix~\ref{append:rpow-posterior} and \eqref{eq:tau-posterior}.  In the following, we discuss their implied hypothesis testing and demonstrate that the robust MAP prior is a more appropriate choice for dynamic borrowing.  

From the Bayesian hypothesis testing perspective, $B_\lambda$ measures the evidence against $H_0: \lambda = 0$ in favor of $H_a^{\pow}: \lambda = \lambda_a$, where $0 < \lambda_a \leq 1$. These two hypotheses can be translated into those testing $1/\tau$, as $H_0: 1/\tau = 0$ and $H_a^{\pow}: 1/\tau = 1/\tau_a$, where $\tau_a$ is determined by $\lambda_a$ (implying a targeted effective sample size) using \eqref{eq:tau2lambda}. The null hypothesis posits \textit{unconditional} independence between $\theta_\ast$ and $\theta_1$. Conversely, the alternative one, as implied by the corresponding hierarchical model $\theta_\ast, \theta_1 \sim \dnorm{\cdot}{\mu}{\tau_a^2}$ (with the distribution for $\mu$ omitted), assumes that $\theta_\ast$ and $\theta_1$ are exchangeable--\textit{conditionally} independent given the common mean $\mu$. This alternative hypothesis is of limited practical relevance, as it is often ill-posed. When $H_0$ is false, the true value of $\tau$ may differ from $\tau_a$ (corresponding to $\lambda_a$ in the power prior). Specifically, setting $\lambda_a = 1$ yields $H_a^{\pow}: \tau = 0$, which forces $\theta_\ast = \theta_1$--an assumption that is typically unrealistic given potential heterogeneity between sources. 

The robust MAP prior effectively addresses this issue. The associated Bayes factor ($B_\nu$) quantifies the evidence against $H_0: \tau^{-1} = 0$ in favor of $ H_a^{\map}: \tau^{-1} > 0$. These are two complementary hypotheses, meaning that one of the two must be true. From a Bayesian model averaging perspective, the robust MAP prior spans the entire model space for  $\theta_\ast$ and $\theta_1$, ranging from unconditional independence to complete equality. In contrast, the robust power prior considers only two discrete points on this continuum, making it overly restrictive.

\section{Application} \label{sec:app}

In this section, we show that our method can improve the efficiency and transparency for the development of Bayesian dynamic borrowing designs in the real-life application introduced in Section~\ref{sec:ex}. The the details for the study design are provided as follows.

A global study with sample size of of 800 ($n_1 = 800$) yields an estimated treatment difference of 86 ($\hy_1 = 86$) and the associated standard error of 20.1 ($\hs_1 = 20.1$). Originally, the Bayesian dynamic borrowing design was planned based on $p_\rpow(\cdot)$ for this bridging study. The parameter of the weakly informative component is set as $\sigma_0 = \sqrt{2} \sigma$, where $\sigma = 350$ represents the assumed standard deviation in the global study design. The power prior for the informative component was specified with $\lambda = 1$. Although the available evidence supported a higher prior weight for the informative component (i.e., $w_0 \geq 0.5$), $w_0$ was set to $0.3$ to maintain the $\alpha$-level at 0.2. The planned sample size of the bridging study was 150 ($n_\ast = 150$) and the success criterion was $\pr ( \theta_\ast > 0 \vert \hy_\ast, \hs^2_\ast, \hy_1, \hs^2_1 ) \geq 95 \%$. The proposed design provided approximately 80\% power to detect a treatment difference of 100, which is the alternative value used in the global study design. 

\subsection{Identifying prior parameters for given operating characteristics}  \label{sec:app-ex1-oc}

In \citet{edwards2024using}, tedious simulations were employed to evaluate operating characteristics, which we argue are unnecessary.  Such calculation can be done in a more efficient manner (costing no more than a few minutes) using the approach described in Section~\ref{rem:pre-bdb-oc} with the closed-form formulas provided in Appendix~\ref{append:rpow-posterior} (for the robust mixture prior) and \eqref{eq:rmap-posterior} (for the robust MAP prior). Appropriate prior specifications can be identified by grid search for both priors under any choice of $w_0$, yielding nearly identical operating characteristics.

Without loss of generality, we fix the prior mixture weight at $w_0 = 0.3, 0.5$ and $0.7$, respectively. For $p_\rpow(\cdot)$, the calibration parameter is $\lambda$. For $p_\rmap(\cdot)$, we use a half-normal distribution for $f_0(\tau;\nu)$, where $\nu$ is the scale parameter under calibration. Besides, we let $n_\ast = 150$, and $\hs_\ast^2 = \sigma^2 / n_\ast$ in \eqref{eq:rmap-posterior} and \eqref{eq:tau-posterior}. Table~\ref{tab:app-oc} summarizes the values of $w_0$, $\lambda$ (if using robust power priors), and $\nu$ (if using robust MAP priors), which lead to the same type I error rate ($\alpha$), power ($1-\beta$) and critical decision boundary ($y_c$). Note that $w_0 = 0.3$ and $\lambda = 1$ correspond to the robust power prior used in \citet{edwards2024using}.

\begin{table}[t!]
\centering
\begin{tabular}{c|ccccc}
\hline
$w_0$ & $\lambda$ & $\nu$ & $\alpha$ & $1-\beta$ & $y_c$ \\ \hline
  0.3 & 1 & 1 & 0.196 & 0.814 & 49 \\
  0.5 & 0.185 & 34 & 0.196 & 0.814 & 49 \\
  0.7 & 0.144 & 46 & 0.196 & 0.814 & 49 \\\hline
\end{tabular}
\caption{The values of parameters in robust power priors ($w_0, \lambda$) and robust MAP priors ($w_0, \nu$) produce the same operating characteristics: $\alpha$ = type I error; $1-\beta$ = power; $y_c$ = critical decision boundary. }
\label{tab:app-oc}
\end{table}

\subsection{Assessing the prior parameter of between-group heterogeneity}  \label{sec:app-ex1-tau}

\begin{figure}[t!]
    \centering
    \includegraphics[width=\linewidth]{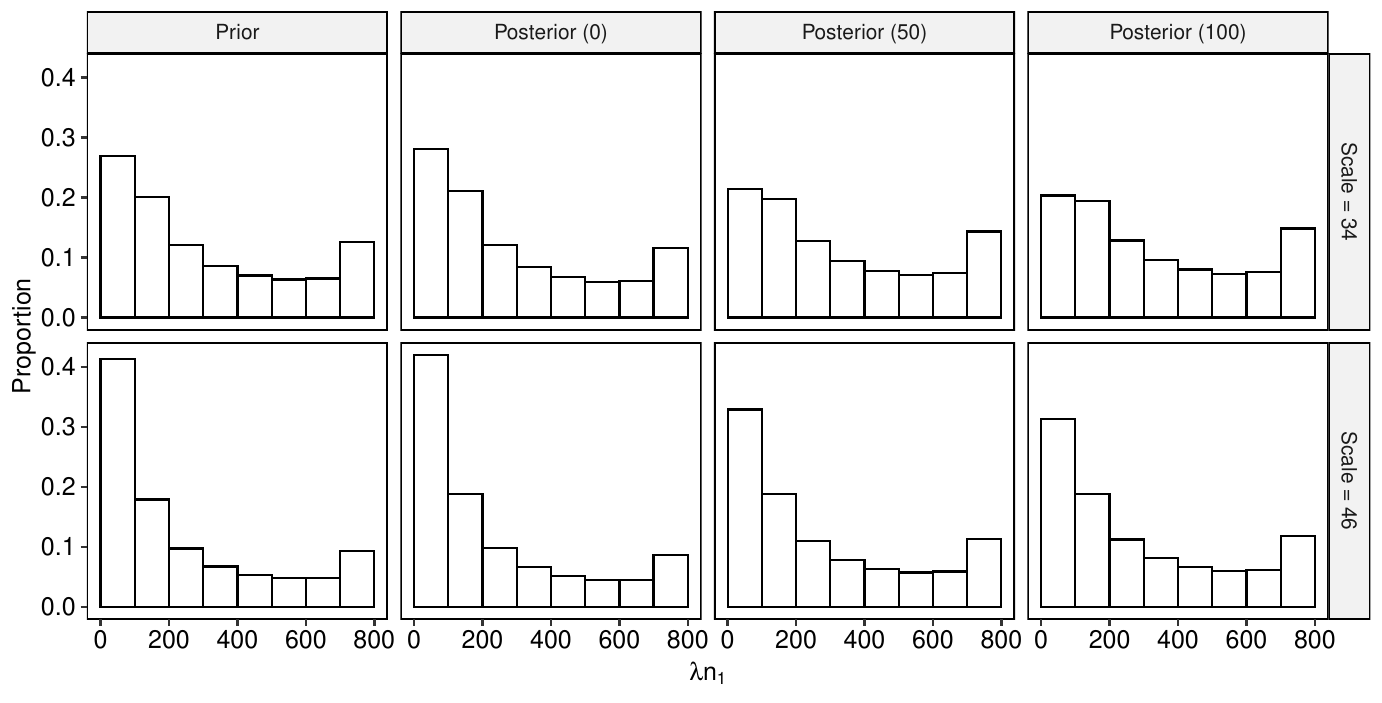}
    \caption{The prior and the posterior distributions (given $\hy_\ast = 0, 50, 100$) of $\lambda n_1$ for the last two MAP priors in Table~\ref{tab:app-oc}. The resulting distributions are summarized into those histograms with the bin width of one hundred. }
    \label{fig:app-tau}
\end{figure}

When using $p_\rmap(\cdot)$, the magnitude of external information incorporated into analysis results through the informative component (i.e., the MAP prior) is presented by each $h(i;\nu)$ in \eqref{eq:tau-posterior}, the posterior probability of $\lambda n_1 = i$. Figure~\ref{fig:app-tau} summaries the prior and the posterior probabilities of $\lambda n_1 = i$ for the two MAP priors with $\nu = 34, 46$ (the last two in Table~\ref{tab:app-oc}), respectively. Those prior probabilities are calculated using $h_{0i}$ in \eqref{eq:tau-prior}. We consider three scenarios of observed data: $\hy_\ast = 0, 50, 100$. Histograms were initially computed with a bin width of $1$ and later aggregated into bins of width $100$ to improve interpretability and facilitate visualization. 

Figure~\ref{fig:app-tau} presents histograms that quantify the contribution of external information to posterior distributions under two MAP priors, each defined by different values of $\nu$ in the half-normal prior for $\tau$. When $\hy_\ast = 0$, the posterior histograms closely resemble the prior distributions, indicating minimal borrowing. In contrast, for $\hy_\ast = 50$ or $100$, the posterior distributions exhibit greater mass at larger $\lambda n_1$ values and reduced mass at smaller ones. The two posterior histograms for $\hy_\ast = 50$ and $100$ are similar, both slightly shifted relative to $\hy_1 = 86$, the mean of the external group. These patterns illustrate how information borrowing varies with $\hy_\ast$. Notably, the posterior histograms at $\hy_\ast = 50$ and $100$ under the MAP prior with $\nu=46$ convey less information than the prior histogram under $\nu=34$, indicating that the former prior is overly conservative.

\section{Discussion} \label{sec:disc}

In this article, we propose a closed-form approximation for computing posterior distributions with robust MAP priors in regional treatment effect assessment.  This estimator greatly facilitates evaluation of operating characteristics in Bayesian dynamic borrowing designs based on robust MAP priors.  In particular, it reduces computation time for type I error and power to just a few minutes, enabling efficient identification of prior parameters.  Using a real-world example, we compare the proposed method with the standard approach based on robust power priors and show that it produces nearly identical designs with minimal additional effort.

In Section~\ref{sec:method-bf}, we show that the null and alternative hypotheses implied by the robust mixture prior are non-complementary, making the corresponding test of whether $\theta_\ast$ and $\theta_1$ are related ill-posed. In contrast, the hypothesis framework based on the robust MAP prior avoids this flaw, suggesting it is an attractive choice for Bayesian dynamic borrowing designs--even though both priors can achieve the same type I error rate and power when prior parameters are properly tuned.

For robust MAP priors, our proposed estimator offers a refined approach for quantifying the effective sample size of the informative component in posterior distributions; see $h(i;\nu)$ in \eqref{eq:tau-posterior} and Figure~\ref{fig:app-tau}. The posterior distribution of $\tau$ can be displayed as a histogram on the sample-size scale after transforming it to $\lambda n_1$, representing the effective sample size derived from a power prior with discounting parameter $\lambda$.  These histograms, capturing uncertainty in the degree of borrowing, provide a more nuanced perspective on the dynamics of information sharing.  

In Section~\ref{sec:app}, we illustrate our method’s utility using a real-world example with continuous outcomes. More generally, the approach is applicable to binary and survival outcomes, provided that treatment effect estimates are asymptotically normal--a condition that typically holds in standard statistical analyses.

\FloatBarrier






 




\bibliographystyle{apalike}        
\bibliography{main}

\clearpage

\appendix
\appendixpage
\beginappendix

\section{Posterior Distributions for Robust Power Priors} \label{append:rpow-posterior}

The robust power prior yields a posterior distribution that is a two-component mixture, expressed as 
\begin{multline*}
\pi_{\rpow} ( \theta_\ast \vert \hy_\ast, \hs_\ast^2, \hy_1, \hs_1^2, \lambda, w_0 ) = \ ( 1 - w ) \cdot \dnorm{\theta_\ast}{\hy_\ast}{ \frac{1}{ \hs_\ast^{-2} + \sigma_\ast^{-2} } } \\ + w \cdot \dnorm{\theta_\ast}{ \frac{ \hs_\ast^{-2} \hy_\ast + \lambda \hs_1^{-2} \hy_1 }{ \hs_\ast^{-2} + \lambda \hs_1^{-2} } }{ \frac{1}{ \hs_\ast^{-2} + \lambda \hs_1^{-2} } },
\end{multline*}
where 
\begin{align*}
w & = \frac{ B_\lambda \cdot w_0 / ( 1 - w_0 ) }{ B_\lambda \cdot w_0 / ( 1 - w_0 ) + 1 } \\
B_\lambda & = \frac{ \dnorm{\hy_\ast}{\hy_1}{ \hs_1^2 / \lambda + \hs_\ast^2 } }{ \dnorm{\hy_\ast}{0}{ \sigma_\ast^2 + \hs_\ast^2 } }.
\end{align*}

\section{Approximating the MAP Prior by a Mixture of Power Priors} \label{append:proof-map-mix-pow}

The MAP prior can be expressed as the predictive distribution based on $\dnorm{\theta_\ast}{\mu}{\tau^2}$ and $\pi ( \mu, \tau \vert \hy_1, \hs_1^2, f_0 ) = \int \! \pi ( \theta_1, \mu, \tau \vert \hy_1, \hs_1^2, f_0 ) \, \mathrm{d} \theta_1$, where 
\[
\pi ( \theta_1, \mu, \tau \vert \hy_1, \hs_1^2, f_0 ) \propto \dnorm{\hy_1}{\theta_1}{\hs_1^2} \dnorm{\theta_1}{\mu}{\tau^2} \dnorm{\mu}{m}{v} f_0 ( \tau; \mu ).
\]
Thus, we have that 
\begin{align*}
& \pi ( \mu, \tau \vert \hy_1, \hs_1^2, f_0 ) \\
\propto & \dnorm{\hy_1}{\mu}{ \hs_1^2 + \tau^2 } \dnorm{\mu}{m}{v} f_0 ( \tau; \nu ) \\
\propto & \underbrace{ \dnorm{\mu}{ \frac{ \hy_1 / ( \hs_1^2 + \tau^2 ) + m / v }{ 1 / ( \hs_1^2 + \tau^2 )+  1 / v } }{ \frac{1}{ 1 / ( \hs_1^2 + \tau^2 ) + 1 / v }} }_{ \propto \pi ( \mu \vert \tau, \hy_1, \hs_1^2 ) } \times \underbrace{ \dnorm{\hy_1}{m}{ \hs_1^2 + \tau^2 + v } f_0 ( \tau; \nu ) }_{ \pi ( \tau \vert \hy_1, \hs_1^2, f_0 ) }.
\end{align*}
Let $m\to0$ and $v\to+\infty$. It immediately indicates that $\pi ( \mu \vert \tau, \hy_1, \hs_1^2 ) = \dnorm{\mu}{\hy_1}{\hs_1^2+\tau^2}$ and $\pi ( \tau \vert \hy_1, \hs_1^2, f_0 ) = f_0 ( \tau; \nu )$, which completes the proof.

\section{Posterior under Approximation of the Robust MAP Prior} \label{append:proof-rmap-posterior-approx}

Given $\dnorm{\hy_\ast}{\theta_\ast}{\hs_\ast^2}$ and \eqref{eq:tau-prior}, the mixture components in the posterior distribution, associated with $\dnorm{\theta_\ast}{0}{\sigma_0^2}$ and each $\dnorm{\theta_\ast}{\hy_1}{n_1/i\cdot\hs_1^2}$, can be obtained using the normal-normal conjugacy as shown in \eqref{eq:rmap-posterior}. Additionally, we have that 
\begin{align*}
 h(i;\nu) & \propto h_0(i;\nu) \int\! \dnorm{\hy_\ast}{\theta_\ast}{\hs_\ast^2} \dnorm{\theta_\ast}{\hy_1}{n_1/i\cdot\hs_1^2} \, \mathrm{d} \theta_\ast \\
 & \propto h_0(i;\nu) \cdot \dnorm{\hy_\ast}{\hy_1}{\hs_\ast^2+n_1/i\cdot\hs_1^2}.
\end{align*}
The Bayes factor $B_\nu$ can be obtained by the ratio of two marginal densities provided below
\begin{align*}
\int \! \dnorm{ \hy_\ast }{ \theta_\ast }{ \hs_\ast^2 } \dnorm{ \theta_\ast }{0}{\sigma_0^2} \, \mathrm{d} \theta_\ast 
& = \dnorm{\hy_\ast}{0}{ \hs_\ast^2 + \sigma_0^2 }, \\
\int \! \dnorm{ \hy_\ast }{ \theta_\ast }{ \hs_\ast^2 } \SUM{i=1}{n_0} h_0(i;\nu) \cdot \dnorm{ \theta_\ast }{ \hy_1 }{ n_1 / i \cdot \hs_1^2 } \, \mathrm{d} \theta_\ast 
& = \SUM{i=1}{n_0} h_0(i;\nu) \cdot \dnorm{ \hy_\ast }{ \hy_1 }{ \hs_\ast^2 + n_1 / i \cdot \hs_1^2 }.
\end{align*}
Following the definition of Bayes factor, the posterior mixture weight $w$ can be obtained directly from $B_\nu$ and $w_0$. The above results complete the proof of \eqref{eq:tau-posterior}.

\section{R Code for the Motivating Example} \label{append:code}

\begin{lstlisting}

# R code to replicate the analysis in Sec 5 

library(dplyr)
library(ggplot2)
library(ggpubr)

# utility functions for robust power priors 

bf_rpp <- function(y, s, y0, s0, sigma0, lambda) {
  return(
    dnorm(y, y0, sqrt(s^2 + s0^2 / lambda)) / 
      dnorm(y, 0, sqrt(s^2 + sigma0^2)))
}

posterior_rpp <- function(y, s, y0, s0, sigma0, lambda, w0) {
  k <- bf_rpp(y, s, y0, s0, sigma0, lambda)
  w <- k * w0 / (1 - w0) / (k * w0 / (1 - w0) + 1)
  prob <- (1 - w) * pnorm(sqrt(1/s^2 + 1/sigma0^2) * y) + w * 
    pnorm((y / s^2 + lambda * y0 / s0^2) / sqrt( 1/s^2 + lambda / s0^2))
  return(prob)
}

# utility functions for robust MAP priors

bf_map <- function(y, s, y0, s0, sigma0, n0, h0) {
  k <- sapply(1:n0, function(i) 
    dnorm(y, y0, sqrt(s^2 + s0^2 / (i / n0))) / 
      dnorm(y, 0, sqrt(s^2 + sigma0^2)))
  return(sum(k * h0))
}

posterior_rmap <- function(y, s, y0, s0, sigma0, n0, h0, w0) {
  k <- bf_map(y, s, y0, s0, sigma0, n0, h0)
  w <- k * w0 / (1 - w0) / (k * w0 / (1 - w0) + 1)
  uh <- h0 * dnorm(y, y0, sqrt(s^2 + s0^2 / (c(1:n0) / n0)))
  h <- uh / sum(uh)
  prob <- (1 - w) * pnorm(sqrt(1/s^2 + 1/sigma0^2) * y) + w * sum(
    h * pnorm((y / s^2 + (c(1:n0) / n0) * y0 / s0^2) / 
        sqrt( 1/s^2 + (c(1:n0) / n0) / s0^2)))
  return(prob)
}

bf_map_hn <- function(y, s, y0, s0, sigma0, n0, scale) {  # for half-normal priors
  h0 <- rep(0, n0)
  h0[1] <- 1 - pnorm(sqrt((n0 - 1) * s0^2 / 2), sd = scale)
  for (i in 2:n0) {
    h0[i] <- pnorm(sqrt((n0 / (i - 1) - 1) * s0^2 / 2), sd = scale) - 
      pnorm(sqrt((n0 / i - 1) * s0^2 / 2), sd = scale)
  }
  return(bf_map(y, s, y0, s0, sigma0, n0, 2*h0))
}

posterior_rmap_hn <- function(y, s, y0, s0, sigma0, n0, scale, w0) {
  # for half-normal priors
  h0 <- rep(0, n0)
  h0[1] <- 1 - pnorm(sqrt((n0 - 1) * s0^2 / 2), sd = scale)
  for (i in 2:n0) {
    h0[i] <- pnorm(sqrt((n0 / (i - 1) - 1) * s0^2 / 2), sd = scale) - 
      pnorm(sqrt((n0 / i - 1) * s0^2 / 2), sd = scale)
  }
  return(posterior_rmap(y, s, y0, s0, sigma0, n0, 2*h0, w0))
}


# Code to generate Table 1

df_rpp <- expand.grid(  # robust power priors
  n = 150,
  y = 0:100,
  n0 = 800,
  y0 = 86,
  s0 = 20.1,
  sigma = 350,
  lambda = (1:800)/800,
  w0 = c(0.3, 0.5, 0.7)
) %>% mutate(
  s = 2 * sigma / sqrt(n),
  sigma0 = sqrt(2) * sigma,
  tau = sqrt((1 / lambda - 1) * s0^2 / 2),
  bf = mapply(bf_rpp, y, s, y0, s0, sigma0, lambda),
  w = bf * w0 / (1 - w0) / (bf * w0 / (1 - w0) + 1),
  pi = mapply(posterior_rpp, y, s, y0, s0, sigma0, lambda, w0))

# operating characteristics 
oc_rpp <- df_rpp %>% filter(pi >= 0.95) %>% group_by(lambda, w0) %>%  
  summarise(bound = min(y), pi = min(pi), w = min(w), n = n[1], sigma = sigma[1]) %>% 
  mutate(alpha = 1 - pnorm(sqrt(n) * (bound - 0) / (2 * sigma)), 
         pow = 1 - pnorm(sqrt(n) * (bound - 100) / (2 * sigma)))

oc_rpp %>% filter(alpha < 0.2 & alpha >= 0.195) %>% group_by(w0) %>% 
  slice_max(lambda, n = 1, with_ties = FALSE) %>% ungroup()

start <- Sys.time()

df_rmap <- expand.grid(  # robust MAP priors 
  n = 150,
  y = 0:100,
  n0 = 800,
  y0 = 86,
  s0 = 20.1,
  sigma = 350,
  scale = 1:60,
  w0 = c(0.3, 0.5, 0.7)
) %>% mutate(
  s = 2 * sigma / sqrt(n),
  sigma0 = sqrt(2) * sigma,
  bf = mapply(bf_map_hn, y, s, y0, s0, sigma0, n0, scale),
  w = bf * w0 / (1 - w0) / (bf * w0 / (1 - w0) + 1),
  pi = mapply(posterior_rmap_hn, y, s, y0, s0, sigma0, n0, scale, w0))

oc_rmap <- df_rmap %>% filter(pi > 0.95) %>% group_by(scale, w0) %>% 
  summarise(bound = min(y), pi = min(pi), w = min(w), n = n[1], sigma = sigma[1]) %>% 
  mutate(alpha = 1 - pnorm(sqrt(n) * (bound - 0) / (2 * sigma)), 
         pow = 1 - pnorm(sqrt(n) * (bound - 100) / (2 * sigma)))

oc_rmap %>% filter(alpha < 0.2 & alpha > 0.195) %>% group_by(w0) %>% 
  slice_min(scale, n = 1, with_ties = FALSE) %>% ungroup()

end <- Sys.time()

end - start # computation time 

# Code to generate Figure 1
 
calc_h0 <- function(i, n0, s0, scale) {
  f_i <- pnorm(sqrt((n0 / i - 1) * s0^2 / 2), sd = scale) 
  f_i_minus_1 <- if_else(i == 1, 1, pnorm(sqrt((n0 / (i-1) - 1) * s0^2 / 2), sd = scale))
  return(2*(f_i_minus_1 - f_i))
}

df_map_tau_p <- expand.grid( 
  n = 150,
  y = NA,
  n0 = 800,
  y0 = 86,
  s0 = 20.1,
  sigma = 350,
  scale = c(34, 46),
  pess = 1:800,
  type = "Prior"
) %>% mutate(prob = calc_h0(pess, n0, s0, scale))

df_map_tau_pi <- expand.grid(  
  n = 150,
  y = c(0, 50, 100),
  n0 = 800,
  y0 = 86,
  s0 = 20.1,
  sigma = 350,
  scale = c(34, 46),
  pess = 1:800
) %>% mutate(
  type = paste0("Posterior (", y, ")"),
  s = 2 * sigma / sqrt(n),
  prior = calc_h0(pess, n0, s0, scale), 
  uposterior = prior * dnorm(y, y0, sqrt(s^2 + s0^2 / (pess / n0)))) %>% 
  group_by(y, scale) %>% mutate(prob = uposterior / sum(uposterior)) %>% 
  ungroup() %>% select(-c(s, prior, uposterior)) 

df_map_tau <- rbind(df_map_tau_p, df_map_tau_pi)

ggplot(df_map_tau) + 
  geom_histogram(aes(x = pess, weight = prob), breaks = 0:8 * 100, closed = "right", 
                 color = "black", fill = "white") +
  labs(x = expression(lambda*n[1]), y = "Proportion") + 
  facet_grid(cols = vars(type), rows = vars(paste0("Scale = ", scale))) + 
  theme_pubr(border = TRUE, legend = "top")

\end{lstlisting}

\end{document}